\documentclass[
    ,final            
  ]
  {aipproc}

\layoutstyle{6x9}

\begin{document}

\title{Identifying Breaks and Curvature in the \textit{Fermi} Spectra of Bright FSRQs and Constraining the Emission Region}


\author{Jonathan Harris}{
  address={University of Durham, Department of Physics, South Road, Durham DH1 3LE, U.K.}
}

\author{Michael K. Daniel}{address={University of Durham, Department of Physics, South Road, Durham DH1 3LE, U.K.}
}

\author{Paula M. Chadwick}{
}

\begin{abstract}
Deviation of the gamma-ray energy spectra of Flat Spectrum Radio Quasars (FSRQs) from a simple power law has been previously observed but the cause of this remains unidentified.  If the gamma-ray emission region is close to the central black hole then absorption of gamma-rays with photons from the broad line region is predicted to produce two spectral breaks in the gamma-ray spectra at fixed energies.  We examine $9$ bright FSRQs for evidence of breaks and curvature.  Although we confirm deviation from a simple power law, break energies are usually not where predicted by the double-absorber model.  In some objects a log-parabola fit is better than a broken power law.  By splitting the data into two equal time epochs we find that the spectral shape of many objects varies over time.
\end{abstract}

\maketitle

\section{Introduction}

Gamma-ray emission has been detected from hundreds of AGN jets by the \textit{Fermi}-LAT and other instruments but still the emission region and emission mechanism remain uncertain. By considering the timescale for observed gamma-ray variability and  the corresponding light crossing times the emission region is estimated to be less than $10^{-3}\rm~pc$ across~\cite{s2011}. The most popular theory for the gamma-ray emission mechanism involves Compton upscattering of lower energy photons either emitted by synchrotron processes within the jet or originating external to the jet. It seems probable that the external Compton component will be more important in Flat Spectrum Radio Quasars (FSRQs) than in BL Lacs since the former have significant numbers of photons from the broad line region (BLR) available as targets for scattering.

The gamma-ray spectra of the FSRQ subclass of AGN deviate from a simple power law description~\cite{abdo2010}.  For 3C 454.3, the brightest FSRQ source in the \textit{Fermi} energy range ($\sim100\rm~MeV$~-~$300\rm~GeV$), a broken power law describes the data better than a simple power law or a log-parabola.  To explain this result a mechanism is required that will produce a sharp break in a spectrum as opposed to a gradual curvature.  In \cite{poutanen2010}, hereafter PS10, the double-absorber model was put forward as an explanation of these spectral breaks.  This model suggests that recombination of Helium and Hydrogen atoms in the BLR creates two increases in the density of low energy photons that in turn cause an increase in optical depth to gamma-rays above certain energies.  These two increases in optical depth occur between $4$~-~$7\rm~GeV$ and $19.2$~-~$30\rm~GeV$ in the object rest frame due to Helium and Hydrogen recombination respectively.  A further approximation puts the increases at $4\rm~GeV$ and $19.2\rm~GeV$. The authors of PS10 show that such increases in optical depth at a given energy can be well approximated by a break in the spectrum at that energy. Therefore, if the gamma-ray emission region lies within the BLR two breaks will occur; the first of these is hypothesized to explain the break seen in~\cite{abdo2010}.

We have examined three years of \textit{Fermi} data for $9$ bright FSRQs to test whether the objects are best described by a simple power law, broken power law or log-parabola.  We also divide the data for each object into two equal time epochs in order to see if the spectral fits and parameters vary with time.  The analysis of PS10 was unable to use the standard \textit{Fermi} analysis software since it did not contain a suitable spectral model.  The analysis used instead made several approximations in the instrument response function and the gamma-ray background that degraded the precision of the fits.  In this work, using a larger data set, we divide the photons from each object into a high and low energy set and determine a break energy for each.  These two breaks were then tested to see if they were in agreement with the predictions of the double-absorber model.  For further details see~\cite{harris}.

\section{Data Analysis}
Following a standard \textit{Fermi} analysis using the P6\_V3\_DIFFUSE instrument response function, each source was modelled as a simple power law, a broken power law and a log-parabola.  The likelihood of each of these fits was used to determine the Akaike Information Criterion (AIC) of each model.  The AIC test is an estimator of the relative Kullback-Leibler information quantities (how much each model diverges from the true distribution of the data compared to the other models, or alternatively how much information is lost by describing it with a particular model~\cite{burnham}).  The AIC test is useful for multi-model testing such as this as it accounts for both the bias introduced by a model having fewer free parameters and the random error introduced by a model having more~\cite{bozdogan}. Differences in AIC of $2$ or more are considered significant (see~\cite{lewis} and references therein).

To test for variation, we split the data from each object into two equal time epochs.  We then repeated the AIC test of the three different spectral models for each time epoch.  We also compared the most likely break energies found for each object in each epoch to see if they are consistent with one another or if variation can be seen when assuming a broken power law description.

The double-absorber model predicts break energies at or just after $4$ and $19.2\rm~GeV$ in the object rest frame.  To test this, we took all of the data below the midpoint of these two energies to be a low energy set, determined the most likely break energy to describe the data and compared it to a simple power law as a null hypothesis.  We then tested the break energies predicted by the double-absorber model to see if they were in agreement with the most likely break energy. We then took all photons above the most likely break energy to form a high energy set and repeated the procedure used on the low energy set.

\section{Results}

\begin{table}
\begin{tabular}{llll}
\hline
  & \tablehead{1}{c}{b}{No Time Cut   }
  & \tablehead{1}{r}{b}{Epoch 1   }
  & \tablehead{1}{r}{b}{Epoch 2   }   \\
\hline
3C~454.3 & BPL* & BPL* & L-P    \\
PKS~1502+106 & L-P* & L-P & BPL*  \\
3C~279 & L-P* & L-P* & L-P*  \\
PKS~1510-08 & BPL* & BPL & BPL  \\
3C~273 & BPL & L-P* & BPL* \\
PKS~0454-234 & L-P & L-P & BPL* \\
PKS~2022-07 & BPL & BPL & BPL \\
TXS~1520+319 & BPL* & L-P & BPL* \\
RGB~J0920+446 & L-P* & L-P & L-P \\
\hline
\end{tabular}
\caption{Best fitting spectral shape for each object with no time cut and in each epoch. BPL is broken power law and L-P is log-parabola. An asterisk indicates that the difference between the fits is significant as determined by an AIC test.}
\end{table}

The best fitting spectral shape for each object with no time cut and in each epoch are shown in Table~1.  All of the sources show significant deviation from a simple power law description.  In five of the nine sources, the best spectral fit changes from a broken power law to a log-parabola or vice versa between epochs.  When comparing the break energies between the two epochs, in six sources the break changes by more than one standard deviation and in three sources the break changes by more than two standard deviations.  Although caution must always be used for a small sample size such as this, the change in break energies seen is more than would be expected from statistical variation alone.

A broken power law was fitted to the low and high energy sets of each object.  In Table~2 we show the exclusion confidence for a simple power law null hypothesis and power laws with break energies predicted by the double-absorber model.  In the low energy set, a simple power law was rejected to $>99\%$ significance in five objects, but in most of these objects, the break energy is in disagreement with those predicted by the double-absorber model.  In the high energy sets, only three objects reject the simple power law null-hypothesis to more than $>99\%$ significance.  Again, in these three objects the break energies found are in disagreement with those predicted by the double-absorber model.

\begin{table}
\begin{tabular}{lccc|ccc}
\hline
  & \tablehead{1}{c}{b}{Simple PL}
  & \tablehead{1}{r}{b}{Low En \\ $4\rm~GeV$}
  & \tablehead{1}{r}{b}{$4$~-~$7\rm~GeV$}
  & \tablehead{1}{r}{b}{Simple PL}
  & \tablehead{1}{r}{b}{High En \\ $19.2\rm~GeV$}
  & \tablehead{1}{r}{b}{$19$~-~$30\rm~GeV$}   \\
\hline
3C~454.3 & $>$99\% & $>$99\% & $>$99\% & $>$99\% & $>$99\% & $>$99\% \\
PKS~1502+106 & 90\% & 86\% & 93\% & $>$99\% & 94\% & $>$99\% \\
3C~279 & $>$99\% & $>$99\% & $>$99\% & 98\% & 94\% & 95\% \\
PKS~1510-08 & $>$99\% & 76\% & $<$1\% & 60\% & 40\% & 53\%  \\
3C~273 & $>$99\% & $>$99\% & $>$99\% & 52\% & 34\% & 47\% \\
PKS~0454-234 & $>$99\% & $>$99\% & $>$99\% & 87\% & 93\% & 99\% \\
PKS~2022-07 & 83\% & 42\% & $<$1\% & 90\% & 83\% & 11\% \\
TXS~1520+319 & $>$99\% & 45\% & $<$1\% & 47\% & 69\% & 63\% \\
RGB~J0920+446 & $>$99\% & $>$99\% & $>$99\% & $>$99\% & 53\% & 70\%\\
\hline
\end{tabular}
\caption{How significantly models can be excluded for the low energy or high energy set of each object.  See text for a full description.}
\end{table}

\section{Conclusions}
All $9$ of the bright FSRQs examined in this work show significant deviation from a simple power law description.  For some objects the best description is a broken power law and for others it is a log-parabola.  Since the log-parabola has one fewer free parameter (we fixed the normalization energy of the log-parabola when fitting) it might be thought that the log-parabola description would be better for the fainter objects in the sample where data is limited.  However, even the second brightest object in our sample, 3C~279, is best described by a log-parabola suggesting that there is a physical difference between sources that are best described by a broken power law and those best described by a log-parabola.

When we fit a broken power law to each epoch, we see weak evidence that the break energy is changing with time.  If the cause of the break is a large-scale feature of the AGN that is stable over long time periods (e.g. the size of the BLR) then variation in the break energy would not be expected.  This is the case with the double-absorber model where the break energies are fixed by the wavelengths of recombination line photons.  Alternatively, if the cause of the break is due to time-varying parameters in the emission region (e.g. the spectrum of particle injection) then variation in the break energy would be expected.

The best fitting spectral shape for an object often changes between epochs.  The reason for this is not clear, it could be a change in the shape of the underlying distribution of particles producing the gamma-rays or it could be that when a spectral shape with time-varying parameters is time-averaged the best description could be a different shape (for example if the spectrum was a broken power law with a break energy that changes with time a log-parabola might be the best description of the time-averaged data).

Our results are in disagreement with the predictions of the double-absorber model published in PS10.  As well as the indication of time variation in the break energy mentioned above, the break energies that we find in the low and high energy sets of the objects are not in agreement with the break energies predicted by the double-absorber model. This work has been funded by the UK STFC.



\bibliographystyle{aipproc}   

\bibliography{sample}



\end{document}